\pdfoutput=1
\documentclass[%
 reprint,
 amsmath,amssymb,
 aps,
 floatfix,
 letterpaper,
 english
]{revtex4-1}

\usepackage{graphicx}% Include figure files
\usepackage{subfig}
\usepackage{floatrow}
\usepackage{dcolumn}% Align table columns on decimal point
\usepackage{bm}% bold math
\usepackage[skip=20pt]{caption}
\usepackage{hyperref}

\usepackage{bxpapersize}% Formating the size of the output dvi file
\makeatletter
\renewcommand\@evenhead{}%
\renewcommand\@oddhead{}%
\renewcommand\@oddfoot{\hfil\thepage\checkindate\hfil}%
\makeatother

\begin{document}

\preprint{APS/123-QED}

\title{Asymmetric Influence of Employees and Trading Partners on Company's Sales\\
and its Dynamical Origin}

\author{Yuh Kobayashi}
 \email{Author Email yuh.kob.2010@gmail.com; kobayashi.y.bz @m.titech.ac.jp}
\affiliation{%
Department of Mathematical and Computing Science, School of Computing, Tokyo Institute of Technology,
Yokohama 226-8502, Japan
}%
\author{Hideki Takayasu}
 \altaffiliation[Also at ]{Institute of Innovative Research, Tokyo Institute of Technology, Yokohama 226-8502, Japan.}%Lines break automatically or can be forced with \\
\affiliation{%
Sony Computer Science Laboratories, Tokyo 141-0022, Japan
}%
\author{Shlomo Havlin}
 \altaffiliation[Also at ]{Institute of Innovative Research, Tokyo Institute of Technology, Yokohama 226-8502, Japan.}
\affiliation{%
Department of Physics, Bar-Ilan University, Ramat-Gan 52900, Israel
}%
\author{Misako Takayasu}
 \email{Corresponding Author Email takayasu.m.aa@m.titech.ac.jp}
 \altaffiliation[Also at ]{Department of Mathematical and Computing Science, School of Computing, Tokyo Institute of Technology, Yokohama 226-8502, Japan.}
\affiliation{%
Institute of Innovative Research, Tokyo Institute of Technology, Yokohama 226-8502, Japan
}%

\date{\today}% It is always \today, today,

\begin{abstract}
Growth of business firms or companies has been a subject of intensive research over a century. However, there still remains controversy about the basic mechanisms of their growth. Inspired by previous work on scaling laws in other systems, here we extend the notion of size of firms from a scalar to a vector in order to characterize in more detail the mechanisms of growth and decay of firms. Based on a large scale dataset of Japanese firms covering over two million firms for two decades (1994--2015), we compile the dataset of vectors of three components, namely, annual sales, number of employee and number of trading partners. We find that the number of employees is more influential in determining firm sales compared to the number of trading partners. This asymmetry is validated by regressions of sales against these parameters and the analysis of growth rate correlations. We then explore multi-variate dynamics of firms by elaborating an evolutionary flow diagram of the averaged motion in the three-dimensional vector space. The flow diagram indicates that firms which deviate from the balanced scaling relation tend to return to this relation. We also find that firms with a chance of large sales growth suffer the risk of high disappearance rate. These results could serve for prediction and modeling of firms, and are relevant for theoretical understanding of the general principles governing complex systems.

\begin{description}
% Secondary publications and information retrieval purposes.
\item[PACS numbers]
02.50.Sk, 89.65.Gh, 89.75.Da
% May be entered using the \verb+\pacs{#1}+ command.
% You may use the \texttt{description} environment to structure your abstract;
% use the optional argument of the \verb+\item+ command to give the category of each item. 
\end{description}
\end{abstract}

\pacs{Valid PACS appear here}% PACS, the Physics and Astronomy
\maketitle

%\tableofcontents

\section{\label{sec:intro}Introduction}

Growth of business firms is not only an important issue for business people but it has been attracting attention of academic researchers for more than a century\cite{Marshall1890,Gibrat1931,Penrose1959,Marris1964,Coad2009}. The origin of models of firm growth dates back to the Gibrat's model\cite{Gibrat1931} which is based on an over-simplified assumption that a firm's growth rate is random and independent of any quantity, even of its own size\cite{Coad2009,Stanley1996}. Recently, analyses of big datasets have shown ubiquity of the fat-tailed distributions of firm growth rates\cite{Stanley1996,Amaral1997,Fu2005,Bottazzi2011}, whose distribution width depends on their sizes\cite{Stanley1996,Amaral1997,DeFabritiis2003,Takayasu2014,Ishikawa2017}. Correspondingly, the rates of firm disappearance show negative dependence on the firm sizes\cite{Ishikawa2017}. Theories have highlighted mechanisms or factors as diverse as hierarchical organization\cite{Buldyrev1997}, stochasticity in competition\cite{Bottazzi2003,Alfarano2012}, financial\cite{Gallegati2003} or hiring/firing\cite{Wright2005} behaviors, preferential attachment of firm `units'\cite{Fu2005,Riccaboni2008}, social networks\cite{Mondani2014} and multiple independent components in firms\cite{Takayasu2014,Malcai1999,Sutton2002,Wyart2003}, to explain those empirical facts on size and growth. However, assumptions behind the theoretical models and their implications are rarely tested using empirical data, and this yields difficulties in reaching consensus on appropriate theoretical frameworks.

We propose here that elusiveness of the nature of firm growth comes from the fact that size of a firm is not a simple scalar quantity, as usually assumed, but it has at least three components; (i) monetary size scaled typically by the annual sales, (ii) labor size measured by the number of employee, and (iii) transaction activity size which can be characterized by the number of direct trading firms. These three quantities are mutually dependent\cite{Coad2010,Coad2012} and have been found to follow non-trivial scaling relations represented by power laws of the form\cite{Saito2007,Watanabe2013}, $y \propto x^a$, which is a typical functional form found in general complex systems, such as animal bodies\cite{Kleiber1947,Stahl1965,Schmidt-Nielsen1984,Martin1993,Savage2004}, ecological communities\cite{Arrhenius1921,Rosenzweig1995}, and cities\cite{Bettencourt2007,Samaniego2008,Li2015a}. For instance, the value of exponent $a$ in the scaling of metabolic rate on body mass have been determined to be very close to 3/4 in mammals\cite{Kleiber1947,Schmidt-Nielsen1984,Savage2004} and theoretically related to the minimization of energy consumption in blood pumping\cite{West1997}. Theoretical considerations, in turn, were able to predict dozens of other scaling exponents in natural systems of animal bodies successfully. Similarly, the unique value of $a = 2$ for humans as compared to $a = 3$ for other animals in the scaling of body mass against body length was theoretically accounted for by human bipedalism\cite{Yi2017}. In this manner, studies of scaling relations could serve as the very basis for a deep understanding of the system's underlying principles. However, our knowledge about scaling relations in other systems including business firms are still very limited. As for the firms system, the questions about multi-variate relationships which reflect the specific mechanisms or factors of firm growth, as well as their dynamical stability through economic changes, are still not settled.

Here we focus our study on the multi-variable relations among these three quantities of firms to clarify their growth mechanisms and their stability by analyzing a comprehensive dataset of about 2 million Japanese firms accumulated over more than 20 years since 1994. In the general framework which we apply to the firm system here, we also draw an explicit analogy between the animal body and firm: we compare annual sales to metabolic rate as it is the rate of activity in terms of money instead of energy, and the number of employees and trading partners to the animal body size. Our results represent the following two major findings.

\textbf{1:} When considering annual sales as a function of both number of employees and number of trading partners, the scaling exponent of employee is found to be significantly higher than that of trading partners. This implies that increasing number of employees affects more strongly the sales growth than the increase of number of trading partners. This fact is directly supported by comparing distributions of sales growth rates under the conditions that either of employee or trading partners is increased within a certain ratio range while the other is kept nearly constant.

\textbf{2:} In the evolutionary flow diagram, we find that firms tend to move back towards the average sales depending on their size. This indicates that the scaling relations would be recovered after perturbation. In fact, the average growth rate of sales is less than a unit for a firm with more sales compared to the `average firm' of the same sizes of employees and trading partners, while otherwise the sales growth is more than a unit. As a result, in order to increase the chance of a large positive growth, a firm should try to deviate from the scaling relations by increasing either (or both) of employee or trading partners. We also find that there are some regions outside the scaling line where disappearance rates of firms are significantly higher. Therefore, a firm must often take an increasing lethal risk to make a positive growth more probable.

In summary, we first show that on average sales are affected more by increasing employees compared to increasing trading partners, and next we find that firms on average move towards the surface of predicted sales as a function of the numbers of employees and trading partners. Additionally we observe a mild variation of the scaling exponents in the period of 1994--2015, which seem to be correlated to GDP variation. It is noteworthy that despite the variation of the scaling exponents, statistical properties of distributions around the multi-variate scaling relation are stationary throughout the whole observed period. Our finding of asymmetric multi-variate scaling and evolutionary flow diagram in a vector space can provide a more lucid and better understanding of mechanisms of firm growth. This multi-variate approach might also be relevant for better understanding the growth of complex systems in general.

\section{Results}
\subsection{Data Compilation}

We compile the data used here from an exhaustive dataset that summarizes the description of firms by a major credit reporter in the period of 1994 to 2015 (COSMOS 2 by Teikoku Databank,\ Ltd.) available to us in January, 2017. The dataset contains total of $2.415 \times 10^6$ firms ($1.263 \times 10^6$ in yearly average), and the number of listed firms is increasing with time (Supplementary Text \ref{sec:s1}, Supplementary Fig.~\ref{fig:s1}). We filtered out a small fraction of financial firms or governmental organizations whose sales are defined very differently from other ordinary firms. We also removed a very small number of sales data which were recorded more than 8 years after the publication of the financial statement. Additionally, we excluded the sales data where the end of fiscal year changes, because some of them are not annual sales. Therefore, our final dataset, consists of totally $2.395 \times 10^6$ firms ($1.247 \times 10^6$ in yearly average), primarily concerns manufacturing, construction or wholesale companies. Here, we have the data of annual sales $s$, which is the analogue of metabolic rate in animal bodies, and the number of employees, $\ell$, which could be analogous to the animal's cell count, as well as the data of firms' birth and death.

We construct from the data, a trading network for every year from the list of trading relationships between firms, to obtain the degree $k$ (i.e.\ the number of trading partners) of every firm. The network includes $3.051\times10^6$ trading links per year on average (see Supplementary Text \ref{sec:s1} for detail). This enables us to consider the transaction size of firms in the network of trading partnership\cite{Watanabe2013}.

\begin{figure*}
 \sidesubfloat[]{
  \includegraphics[keepaspectratio, scale=0.55]{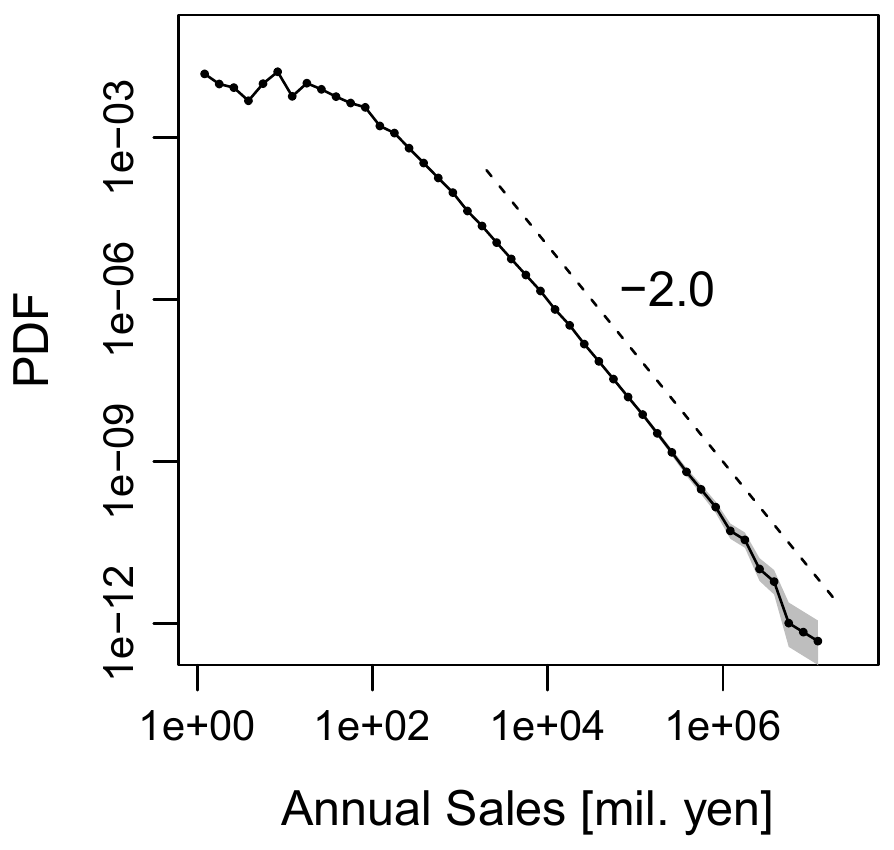}\label{fig:1a}
 }
 \hfil
 \sidesubfloat[]{
  \includegraphics[keepaspectratio, scale=0.55]{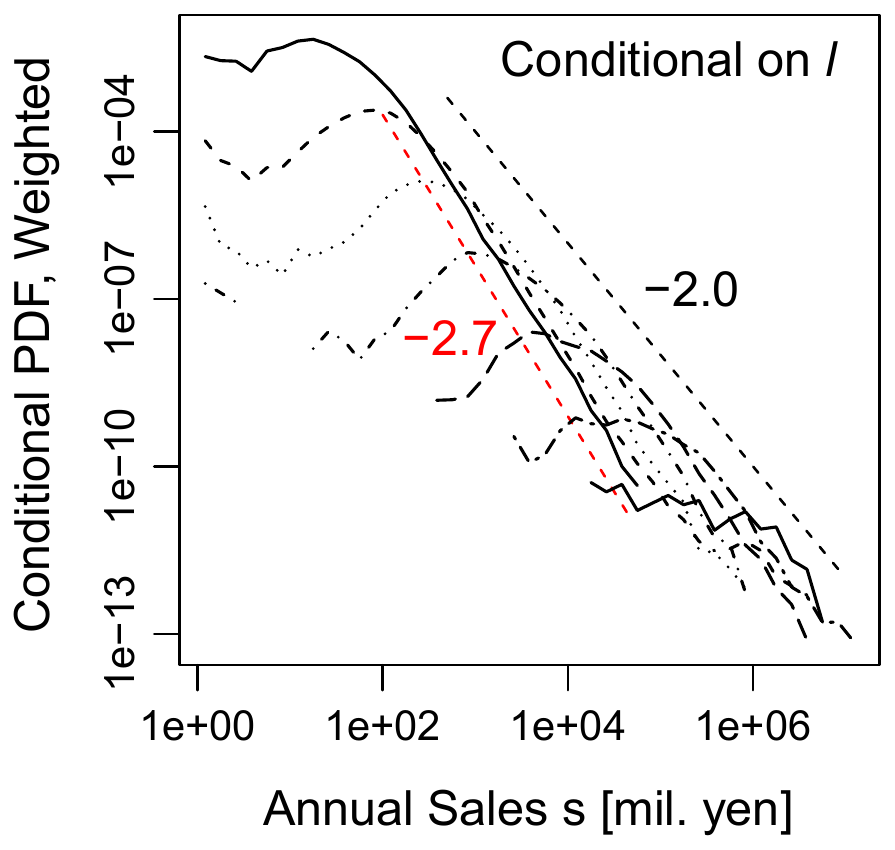}\label{fig:1b}
 }
 \hfil
 \sidesubfloat[]{
  \includegraphics[keepaspectratio, scale=0.55]{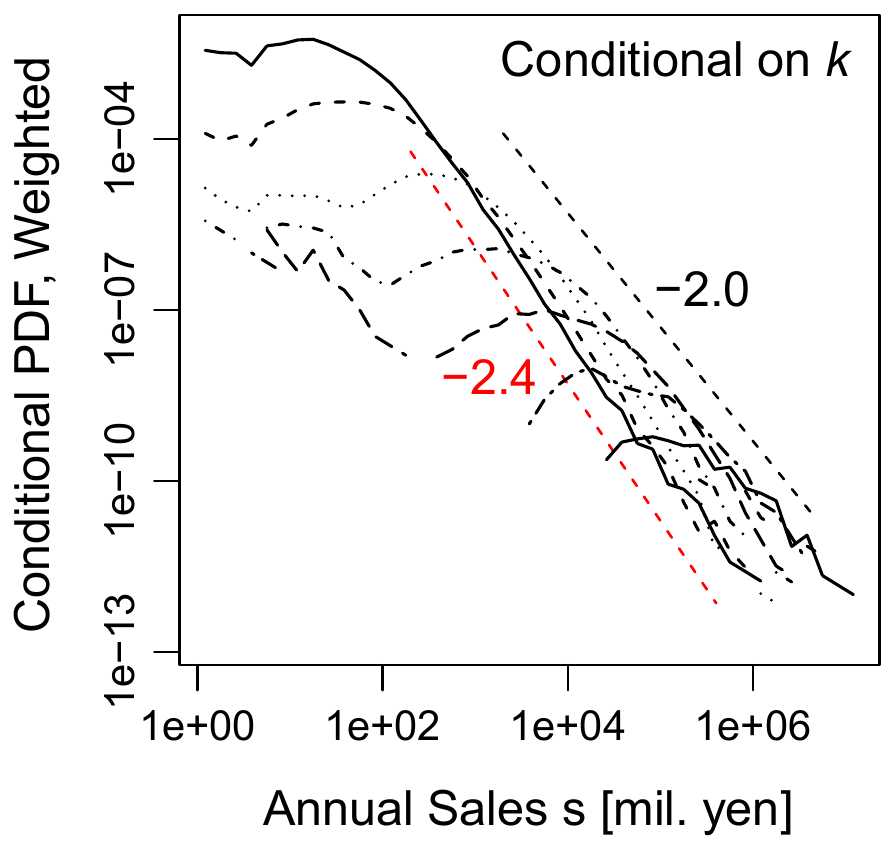}\label{fig:1c}
 }%
 \caption{\label{fig:1} Apparent inconsistency of the power-law tail exponents for marginal and conditional probability distribution functions (PDF), with an intuitive understanding based on the Bayes' theorem. (\textbf{a}) The probability distribution (PDF) of annual sales, plotted in log-log scales, shows a clear exponent of 2. Grey bandwidth indicates 95\% confidence intervals with the assumption of Poisson process. (\textbf{b--c}) The conditional probability distributions ${\rm P}(s|\ell)$ and ${\rm P}(s|k)$, weighted with ${\rm P}(\ell)$ and ${\rm P}(k)$ (see Eq.\ (1)), plotted in log-log scales. The entire range of $k$ and $\ell$ is divided into 8 levels corresponding to 8 curves, so that each interval has an identical range in the logarithmic scale. The weight of an interval is defined as the average probability density in the interval.}
\end{figure*}

We begin by studying the open question of different power law exponents in the conditional and marginal sales distributions, which has not been addressed or mentioned anywhere else to the best of our knowledge. This leads us to the finding of asymmetric role of different aspects of firm body sizes, namely employees, $\ell$, and trading partners, $k$, on the firm sales, $s$.

\subsection{Explanation of Puzzling Scaling Exponents}

It is well established that the distribution of annual sales s of firms roughly follows the Zipf's law\cite{Cirillo2009,Ogwang2011}, that is, the probability density tail follows a power law, ${\rm P}(s) \propto s^{-2}$. This is also seen in our data (see Supplementary Fig.~\ref{fig:s2}c). However, when we look at the conditional sales distributions\cite{Watanabe2013}, the power law exponents increase to about 2.4 (for the number of trading partners $k$---see Fig.~\ref{fig:1c}) and about 2.7 (for employee number $\ell$---see Fig~\ref{fig:1b}) and are clearly different from 2.0 (Fig.~\ref{fig:1a} and Supplementary Figs.~\ref{fig:s4}e and \ref{fig:s4}f). This seemingly contradicting results can be understood by using the Bayes' theorem as follows. Taking the case of the number of employees $\ell$ and annual sales $s$, we can approximate the integral by the contribution of the maximum values of ${\rm P}(s|\ell){\rm P}(\ell)$:
\begin{equation}
{\rm P}(s)
= \int_{\ell} {\rm P}(s|\ell) {\rm P}(\ell) {\rm d}\ell
\sim {\rm P}(s|\ell_{\rm lead}) {\rm P}(\ell_{\rm lead}) {\rm \Delta}\ell_{\rm lead},
\end{equation}
where $\ell_{\rm lead} = {\rm arg\,max}_{\ell}~{\rm P}(s|\ell) {\rm P}(\ell)$, such that ${\rm P}(s|\ell) {\rm P}(\ell)$ is the `leading order' contribution, and ${\rm \Delta}\ell_{\rm lead}$ is the width of $\ell$ at $\ell_{\rm lead}$, which is assumed to be a constant. Indeed, when we plot in Fig.~\ref{fig:1b} the functional form of ${\rm P}(s|\ell) {\rm P}(\ell)$ for several typical values of $\ell$ based on real data, it shows clearly that the envelop function of ${\rm P}(s|\ell) {\rm P}(\ell)$ actually follows a power-law with the exponent close to $-2.0$ at its tail. Similar results are obtained also for the number of trading partners, $k$, as shown in Fig.~\ref{fig:1c}. A more rigorous derivation is given in Supplementary Text \ref{sec:s3}. Thus, Fig.~\ref{fig:1} strongly suggest the origin of the well-known power law exponent of $-2$ for sales distribution\cite{Cirillo2009,Ogwang2011}.

Since the known scaling relations between the size variables\cite{Watanabe2013}, $\ell \propto k^{1.0}$, $s \propto k^{1.2}$ and $s \propto \ell^{1.2}$ (see Supplementary Texts \ref{sec:s2} and \ref{sec:s4}) suggest symmetric roles played by $k$ and $\ell$ in determining the annual sales $s$, it is surprising that our results in Fig.~\ref{fig:1} suggest strong asymmetry between the effects of $k$ and $\ell$. Indeed, in view of magnitude of errors or fluctuations around the scaling relations, the distribution of residuals is more fat-tailed when sales, $s$, is regressed against $k$ rather than when it is regressed against $\ell$. This difference in the tails of fluctuation distributions implies that the number of trading partners, $k$, is less dominant in predicting the sales value compared to the number of employees $\ell$. We discuss this novel feature in more detail in the next section.

\subsection{Multi-variate Scaling}

\begin{figure*}
 %\begin{subfloatrow}[3]
 \sidesubfloat[]{
  \includegraphics[keepaspectratio, scale=0.55]{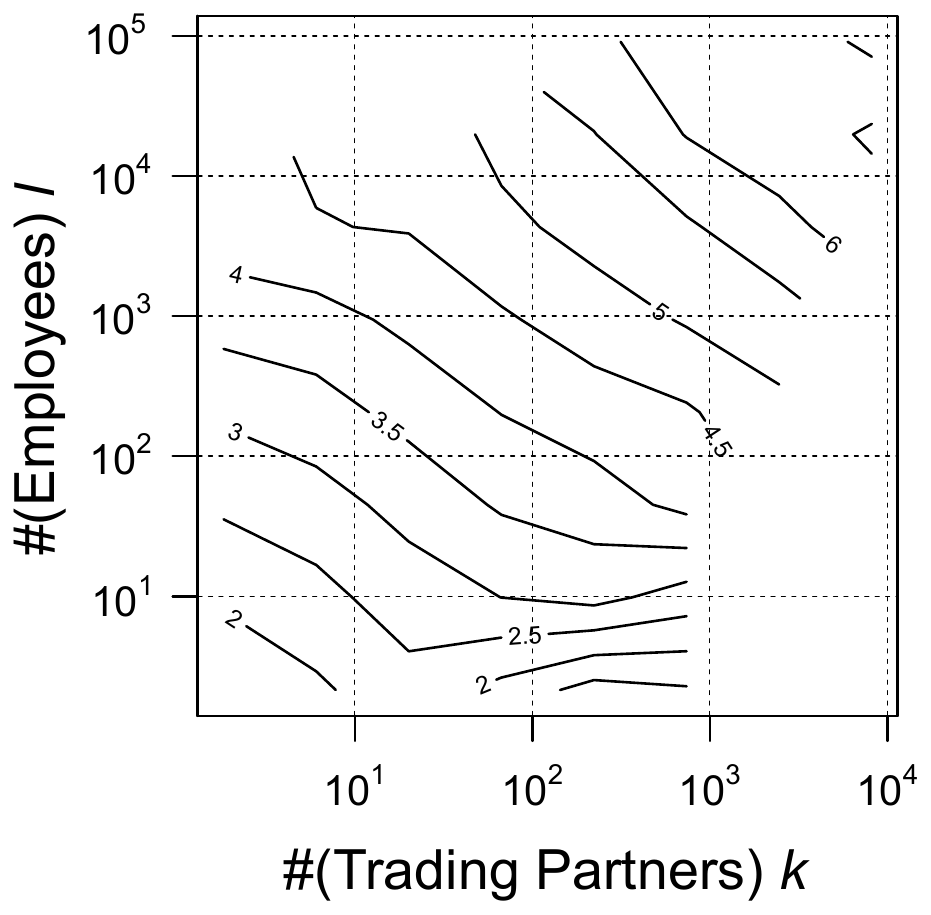}\label{fig:2a}
 }
 \hfil
 \setcounter{subfigure}{2}%
 \sidesubfloat[]{
  \includegraphics[bb=0 0 264 246, keepaspectratio, scale=0.55]{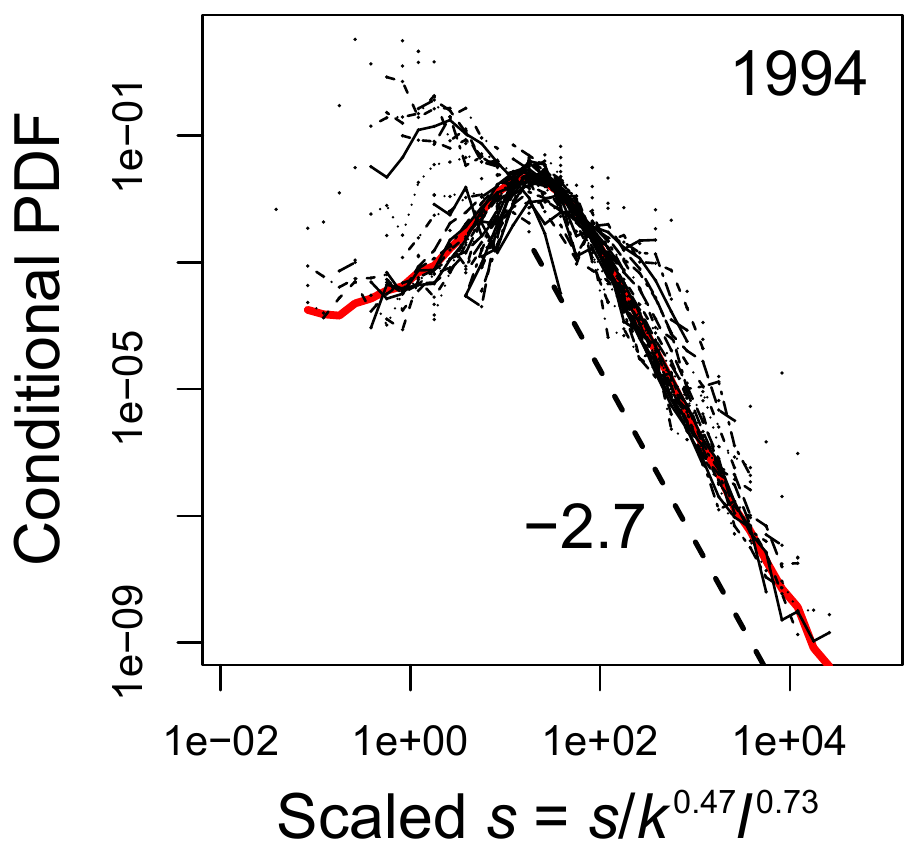}\label{fig:2c}
 }
 \hfil
 \setcounter{subfigure}{4}%
 \sidesubfloat[]{
  \includegraphics[bb=0 0 264 246, keepaspectratio, scale=0.55]{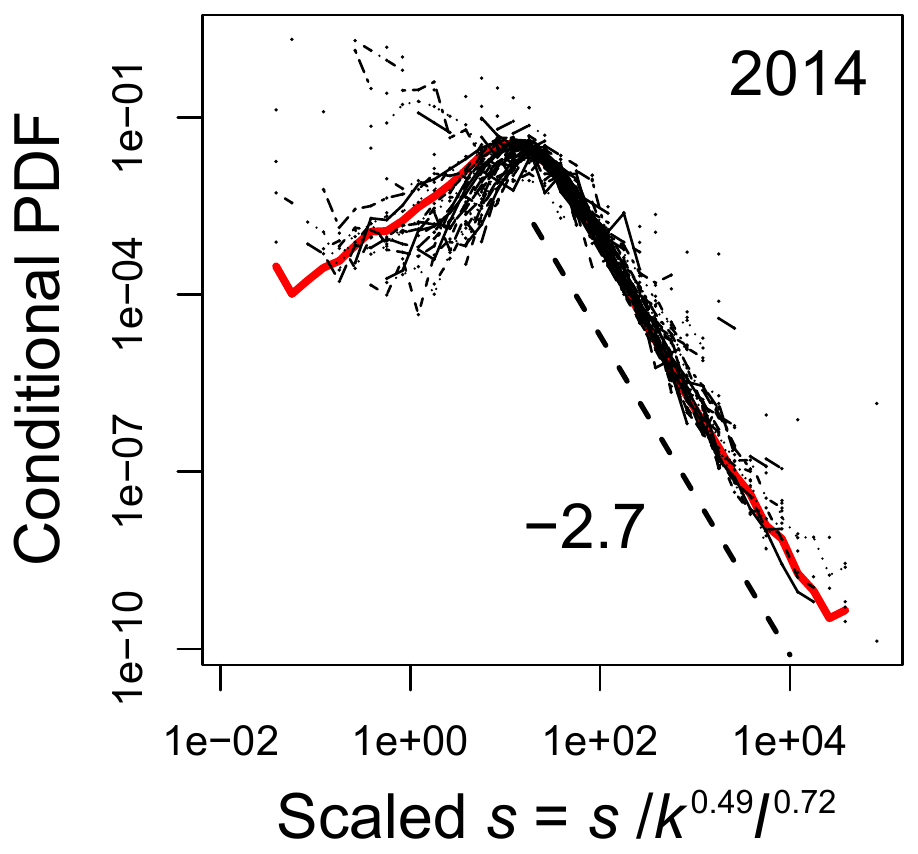}\label{fig:2e}
 }
 \vspace{12pt}
 %\end{subfloatrow}
 %\begin{subfloatrow}[3]
 \setcounter{subfigure}{1}%
 \sidesubfloat[]{
  \includegraphics[keepaspectratio, scale=0.55]{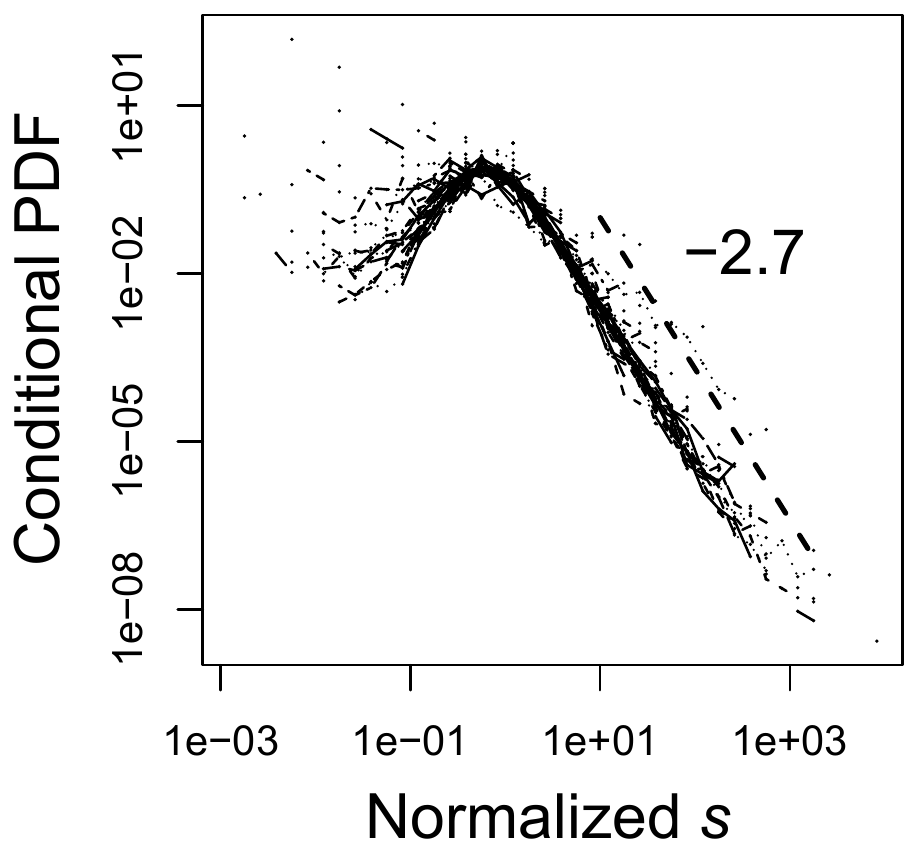}\label{fig:2b}
 }
 \hfil
 \setcounter{subfigure}{3}%
 \sidesubfloat[]{
  \includegraphics[bb=0 0 264 246, keepaspectratio, scale=0.55]{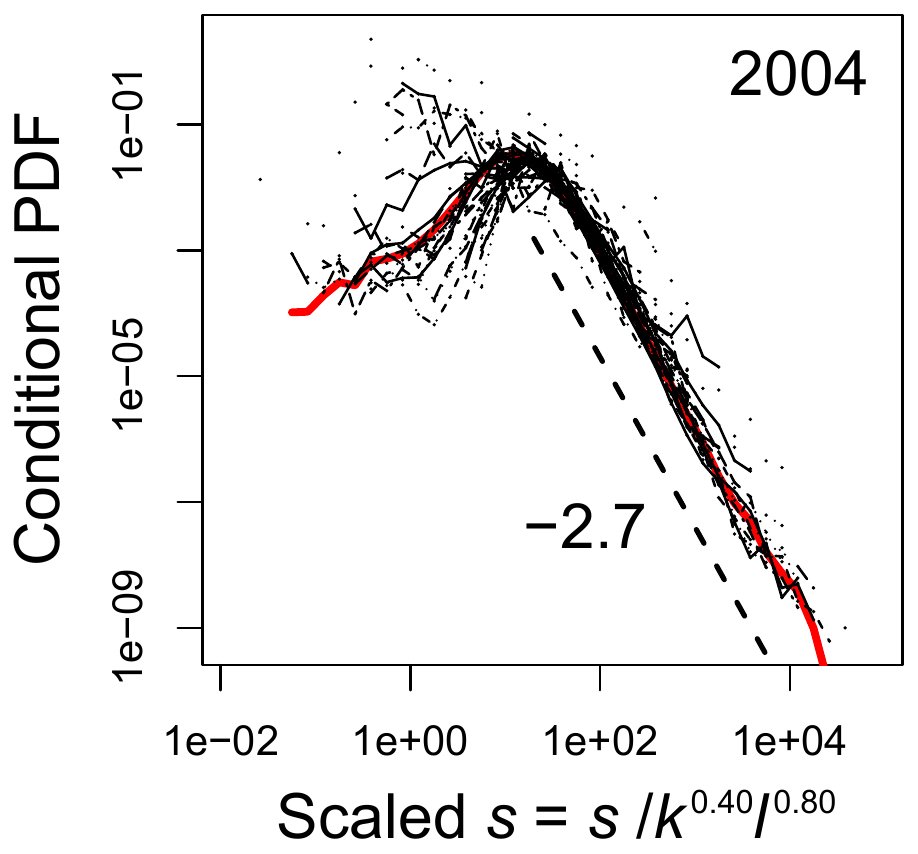}\label{fig:2d}
 }
 \hfil
 \setcounter{subfigure}{5}%
 \sidesubfloat[]{
  \includegraphics[keepaspectratio, scale=0.55]{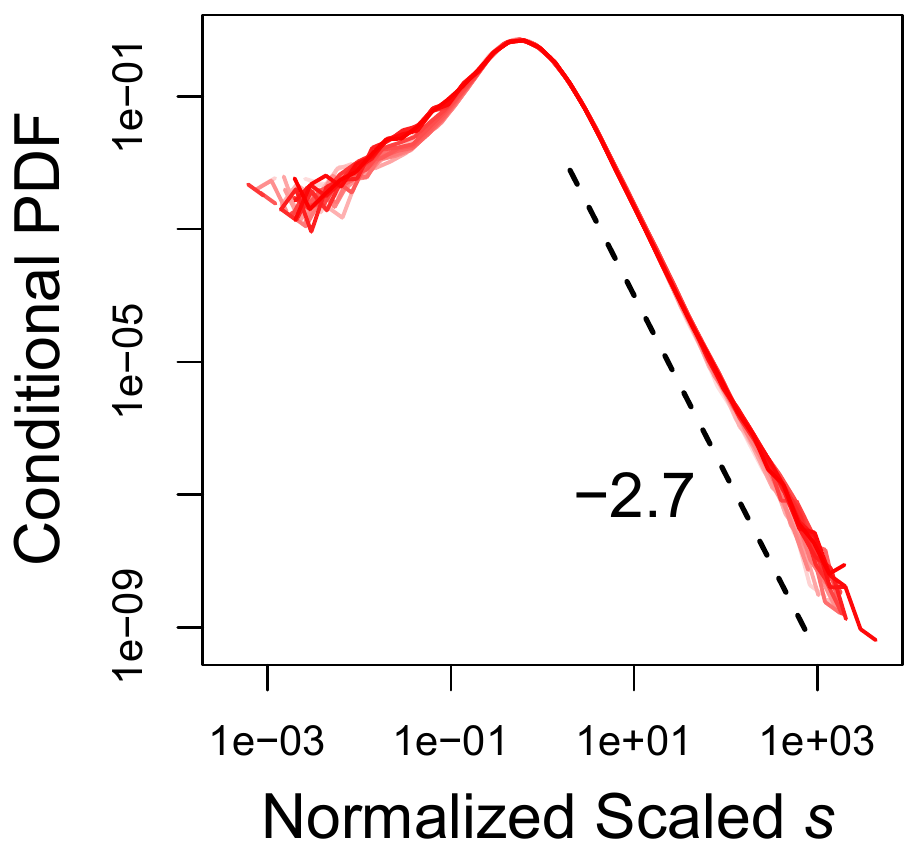}\label{fig:2f}
 }%
 %\end{subfloatrow}
 \caption{\label{fig:2} Multi-variate scaling relations among the three variables. All plots are in a log-log scale. (\textbf{a}) Contour plot of the median values of annual sales $s$ in million yen, conditional on both the number of trading partners $k$ (horizontal) and the number of employees $\ell$ (vertical), for the 2014 data. The entire ranges of $k$ and $\ell$ are divided into 8 levels, so that each interval has an identical range in log scale. The contours are obtained by linearly interpolating the log-transformed median values of sales. (\textbf{b}) Probability distributions (PDF) of $s$ conditional on both $k$ and $\ell$, normalized by their medians, are plotted for the 2014 data. The conditional distributions are obtained for grids of the conditioning variables where both dimensions are divided into intervals of an identical length in log scale (2 segments per a 10-fold interval). (\textbf{c--e}) The probability distributions of scaled sales $s/k^{\alpha}\ell^{\beta}$ conditional on both $k$ and $\ell$, with $\alpha$ and $\beta$ being the estimated exponents. The three panels represent results for 3 different years, namely 1994, 2004 and 2014. Red lines indicate the probability distribution of $s/k^{\alpha}\ell^{\beta}$ without any condition. The method is the same as in panel (b). (\textbf{f}) The probability distributions of $s/k^{\alpha}\ell^{\beta}$ for each of the 22 years (1994--2015), normalized with the medians. The exponents used are different for different years and are taken from the best values found for each year (see also Fig.~\ref{fig:4}).}
\end{figure*}

To assess the relative contributions of the number of trading partners, $k$, and the employee number, $\ell$, on the annual sales $s$, we here generalize the scaling relationship to a multi-variate relation as follows:
\begin{eqnarray}
{\rm log}~s = \alpha~{\rm log}~k + \beta~{\rm log}~\ell + \varepsilon_{s \mid k,\ell}, \\
{\rm P}(s|k,\ell) = \widetilde{P}_{s \mid k,\ell}(s/k^\alpha \ell^\beta)/k^\alpha \ell^\beta,
\end{eqnarray}
\vspace{-30.2pt}\\
\vspace{6.2pt}or\\ %\\[-\baselineskip]
where $\alpha$ and $\beta$ are the scaling exponents indicating the relative effect of $k$ and $l$, $\varepsilon_{s \mid k,\ell}$ a stochastic fluctuation term of ${\rm log}~s$ conditional on both $k$ and $\ell$, ${\rm P}(s|k,\ell)$ is the conditional probability density dependent on both $k$ and $\ell$, and $\widetilde{P}_{s \mid k,\ell}$ is the scaling function. This multiple regression model roughly means $s \propto k^\alpha l^\beta$, and was proposed but not explored nor confirmed by real data in ref.\ \cite{Watanabe2013}. Note that Eq.\ (2) is equivalent to a more formal model of regression against the orthogonalized set of variables, ${\rm log}[k]$ and ${\rm log}[\ell/k^{1.0}]$. Also note that the correlation between growth rates of $k$ and $\ell$ is rather weak (Supplementary Text \ref{sec:s5}, Supplementary Fig.~\ref{fig:s8}). Assuming that Eq.\ (2) is met, it is straightforward to derive the median value as
\begin{equation}
{\log}\langle s|k,\ell \rangle_{0.5} = \alpha~{\rm log}~k + \beta~{\rm log}~\ell + \langle \varepsilon_{s \mid k,\ell} \rangle_{0.5},
\end{equation}
where $\langle \varepsilon_{s \mid k,\ell} \rangle_{0.5}$ is the median value of $s$ conditional on a specific set of $k$ and $\ell$ values. If this is true, the contour plots of conditional median sales $\langle s|k,\ell \rangle_{0.5}$ on the $k-\ell$ logarithmic coordinate plane should show nearly regular and parallel contours. Indeed, from Fig.~\ref{fig:2a} we see that the data actually supports this expectation, especially for the medium or large values. Moreover, we find clearly that the statistical fluctuations around the median value is invariable regardless of the value of $k$ and $\ell$ (Fig.~\ref{fig:2b}). This indicates that the assumption of a scaling function $\widetilde{P}_{s \mid k,\ell}$ is valid for most of $k$ and $\ell$ values. Indeed, when the distribution of scaled sales $s/k^\alpha \ell^\beta$ conditional on $k$ and $\ell$ is plotted using the values of $\alpha$ and $\beta$ estimated based on the data (Figs.~\ref{fig:2c},~\ref{fig:2d} and \ref{fig:2e}), we see that a remarkable fraction of the curves scale with each other. In addition, the function $\widetilde{P}_{s \mid k,\ell}$ is surprisingly stable across years (Fig.~\ref{fig:2f}). Thus, the scaling assumptions of Eqs.\ (2) and (3) are well supported by the large amount of available data.

\begin{figure*}
 {
 \hspace{15mm}
 \sidesubfloat[]{
  \includegraphics[keepaspectratio, scale=0.8]{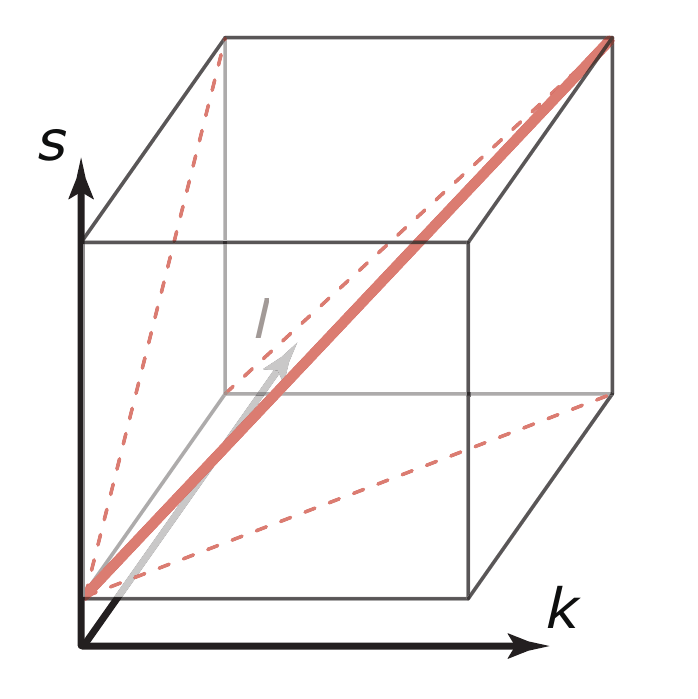}\label{fig:3a}
 }
 \hfil
 \sidesubfloat[]{
  \includegraphics[keepaspectratio, scale=0.8]{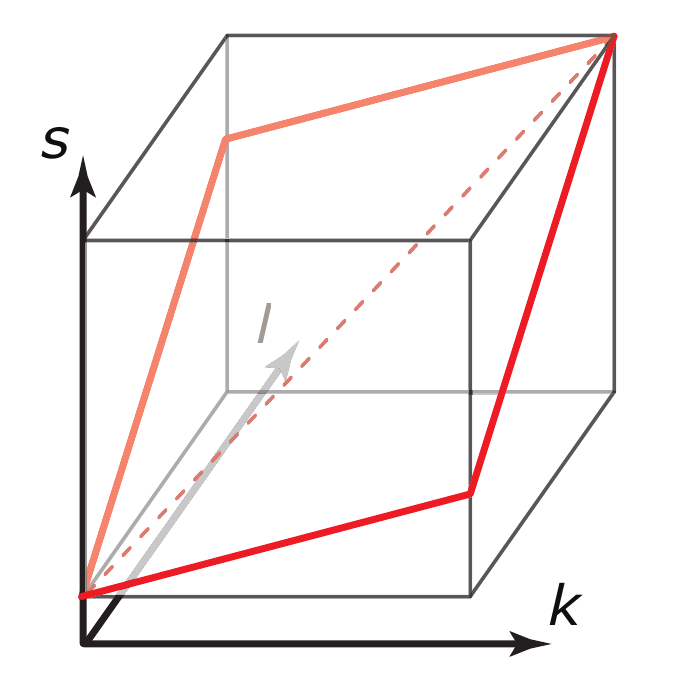}\label{fig:3b}
 }
 \hspace{15mm}
 }%
 \caption{\label{fig:3} Schematic illustrations of the scaling relationships. The variables $k$, $\ell$ and $s$ respectively represent the number of trading partners, employee number and annual sales, and a firm is represented as a point in the 3-dimensional variable space. (\textbf{a}) Three bivariate scaling relations, indicated with the red dashed lines, can be understood as the projections of a single (red bold) `scaling line'. (\textbf{b}) The multi-variate scaling $s \propto k^{\alpha}\ell^{\beta}$ is illustrated as a plane (red solid line). The plane must include the scaling (red dashed) line, but this line clearly cannot determine a unique plane.}
\end{figure*}

\begin{figure}[b]
 {
  \includegraphics[keepaspectratio, scale=0.45]{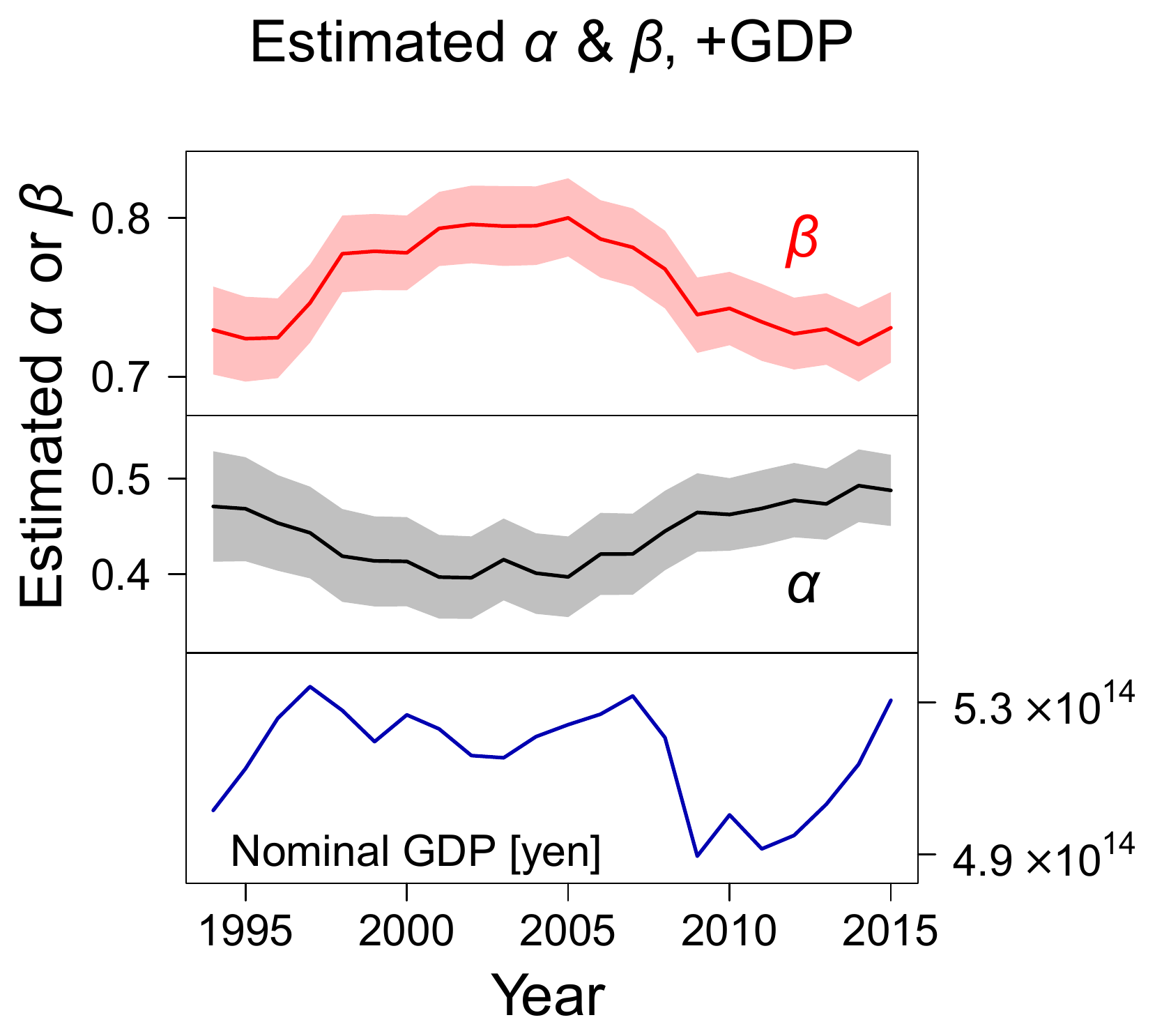}
 }%
 \caption{\label{fig:4} Estimated exponents $\alpha$ (for the number of trading partners $k$) and $\beta$ (for the number of employees $\ell$) compared (in bottom plot) with the country's nominal GDP for different years in 1994--2015. Bandwidth indicates the 95\% confidence interval of the estimation obtained by the bootstrap method.}
\end{figure}

Since $\langle \varepsilon_{s \mid k,\ell} \rangle_{0.5}$ in Eq.\ (4) is a constant, the concept of multi-variate scaling relation can be illustrated by a plane as shown in Fig.~\ref{fig:3b}. In reality, the relation is not a perfect plane, but a surface since it is curved at high-$k$ and low-$\ell$ region as shown in Fig.~\ref{fig:2a}. Importantly, this marks a contrast to a `scaling line' that is implied by the three scaling relationships between pairs of three variables (Fig.~\ref{fig:3a}). In fact, the bivariate scaling laws found in an earlier study\cite{Watanabe2013} are naturally interpreted as projections of a single scaling line to the 2-dimensional planes, where the firms are densely distributed\cite{Klingenberg2016}.

Our finding of $\beta > \alpha$ remains true for all years, as shown in Fig.~\ref{fig:4}, where the estimated scaling exponents for different years are plotted. For example, we have $\alpha = 0.49$ (95\% confidence interval (CI) $(0.455, 0.531)$) and $\beta = 0.72$ (95\% CI $(0.697, 0.744)$) for 2014. Thus, we expect that the employee number $\ell$ actually has a larger effect on sales $s$ compared to the number of trading partners $k$.

Rather unexpected is the gradual change of the values of scaling exponents that seemingly follow the economic climate. Although the inequality $\beta > \alpha$ is maintained, there is a significant change (i.e.\ a change beyond the CI) of $\alpha$ and $\beta$ during the 1994--2015 period, and the changes seem almost in coherence with the nominal GDP, as shown in Fig.~\ref{fig:4} (bottom). Indeed, the cross correlation of estimated $\alpha$ and $\beta$ to the nominal GDP are maximal at the time lag of one year, and as high as $-0.59$ for $\alpha$ and $0.58$ for $\beta$ (Supplementary Text \ref{sec:s4}; Supplementary Fig.~\ref{fig:s6}c). This suggests that the exponents are strongly affected by the GDP of the preceding year. However, the exact causes of this coherence is yet to be undestood.

\subsection{Growth Correlations}

\begin{figure*}
 {
 \hspace{15mm}
 \sidesubfloat[]{
  \includegraphics[bb=0 0 257 242, keepaspectratio, scale=0.66]{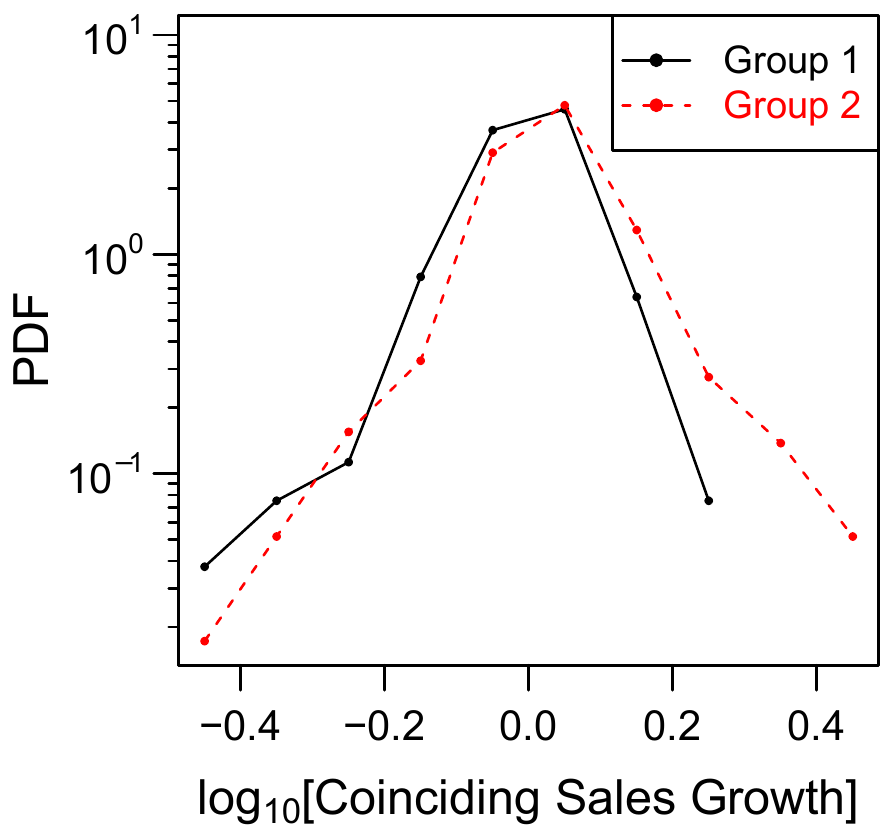}\label{fig:5a}
 }
 \hfil
 \sidesubfloat[]{
  \includegraphics[bb=0 0 257 242, keepaspectratio, scale=0.66]{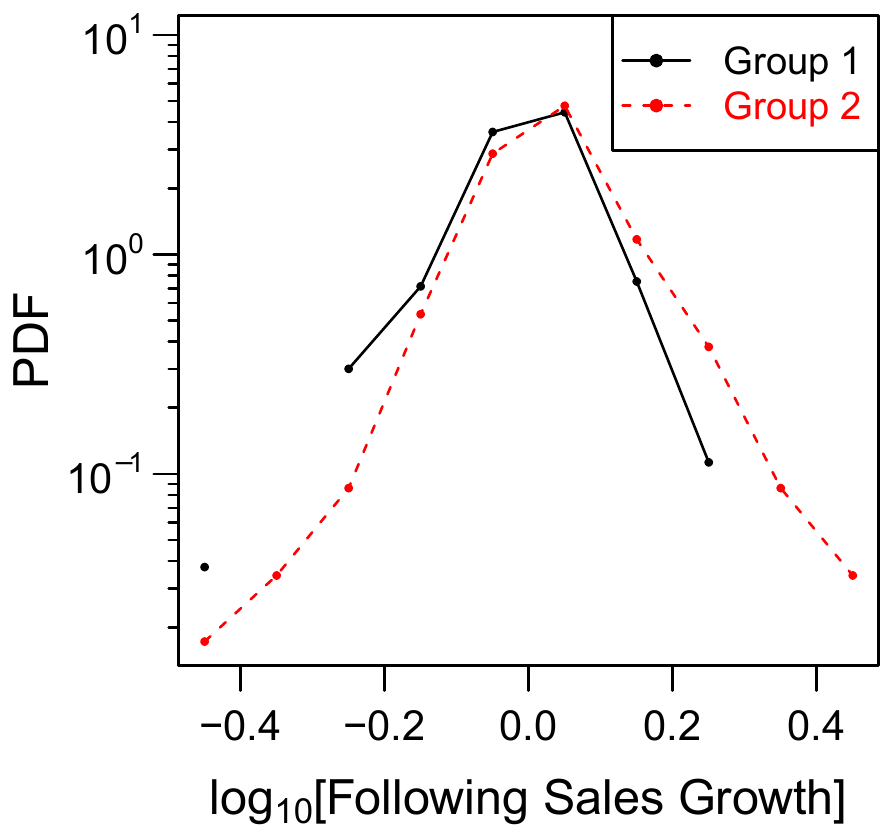}\label{fig:5b}
 }
 \hspace{15mm}
 }%
 \caption{\label{fig:5} Comparison of the effects of increase in trading partners and employees on sales growth. The data is aggregated for 20 years (1994--2013), and selected from around the point of scaling line for $k=10$. The probability distributions (PDF) of log-transformed growth of annual sales, conditional on large growths of $k$ (Group 1, black) and $\ell$ (Group 2, red). The conditions are $1.5 < g_k$ and $0.8 < g_\ell < 1.2$ for large $k$ growth, and $0.8 < g_k < 1.2$ and $1.5 < g_\ell$ for large $\ell$ growth, where $g_k$ and $g_\ell$ respectively denotes the growth rate of $k$ and $\ell$. (\textbf{a}) Sales growth rates at the same year as the size (employees or partners) growth. (\textbf{b}) Sales growth rates in the subsequent year of the size growth.}
\end{figure*}

Next we test, in more detail, the asymmetry between the influence of trading partners and employees on sales. We pay attention to the firms which are on the scaling surface in one year and deviate from it in the following year, and observe their growth in annual sales. In Group 1, we include those firms that increase the number of trading partners by over a factor of 1.5 while keeping their number of employee to be $\pm 20\%$ around the original number. Similarly, in Group 2, we choose firms whose growth rate in employees is over 1.5 while their simultaneous change in the number of trading partners is within $\pm 20\%$ around their original. Fig.~\ref{fig:5a} shows the sales growth distributions for Group 1 (Black) and Group 2 (Red). The probabilities of negative sales growth is generally higher for firms with positive growth in trading partnerships, and higher sales growth is more probable for those with employee growth rather than for those with the same level of growth in trading partnerships. The mean log-transformed growth rate of sales suggests that there is an actual difference: $-0.009$ for firms with growth in trading partnership with 95\% CI (confidence interval) of $(-0.032, 0.014)$, and 0.067 for growth in employees with 95\% CI of $(0.049, 0.086)$. Consistently, two-sample Kolmogorov-Smirnov test also indicates that the difference is significant ($D = 0.164$, $(N_1, N_2) = (266, 581)$, $P \sim 1.2\times10^4$). In the following year, the difference of sales growth rate still remains clear as shown in Fig.~\ref{fig:5b}. Average log-transformed sales growth of Group 1 is 0.000 a year after with 95\% CI of $(-0.027, 0.027)$, while in Group 2 it is 0.068 with 95\% CI of $(0.048, 0.087)$. Also, significance is proved using two-sample Kolmogorov-Smirnov test ($D = 0.151$, $(N_1, N_2) = (266, 581)$, $P \sim 5.1 \times 10^4$). In both cases, the correlation to the sales growth is statistically significant but not strong, indicating that a rapid increase of employee number does not guarantee an immediate growth of sales but only increase the chance.

\subsection{Evolutionary Phase Diagrams}

\begin{figure*}[p]
 {
  \includegraphics[keepaspectratio, scale=0.41]{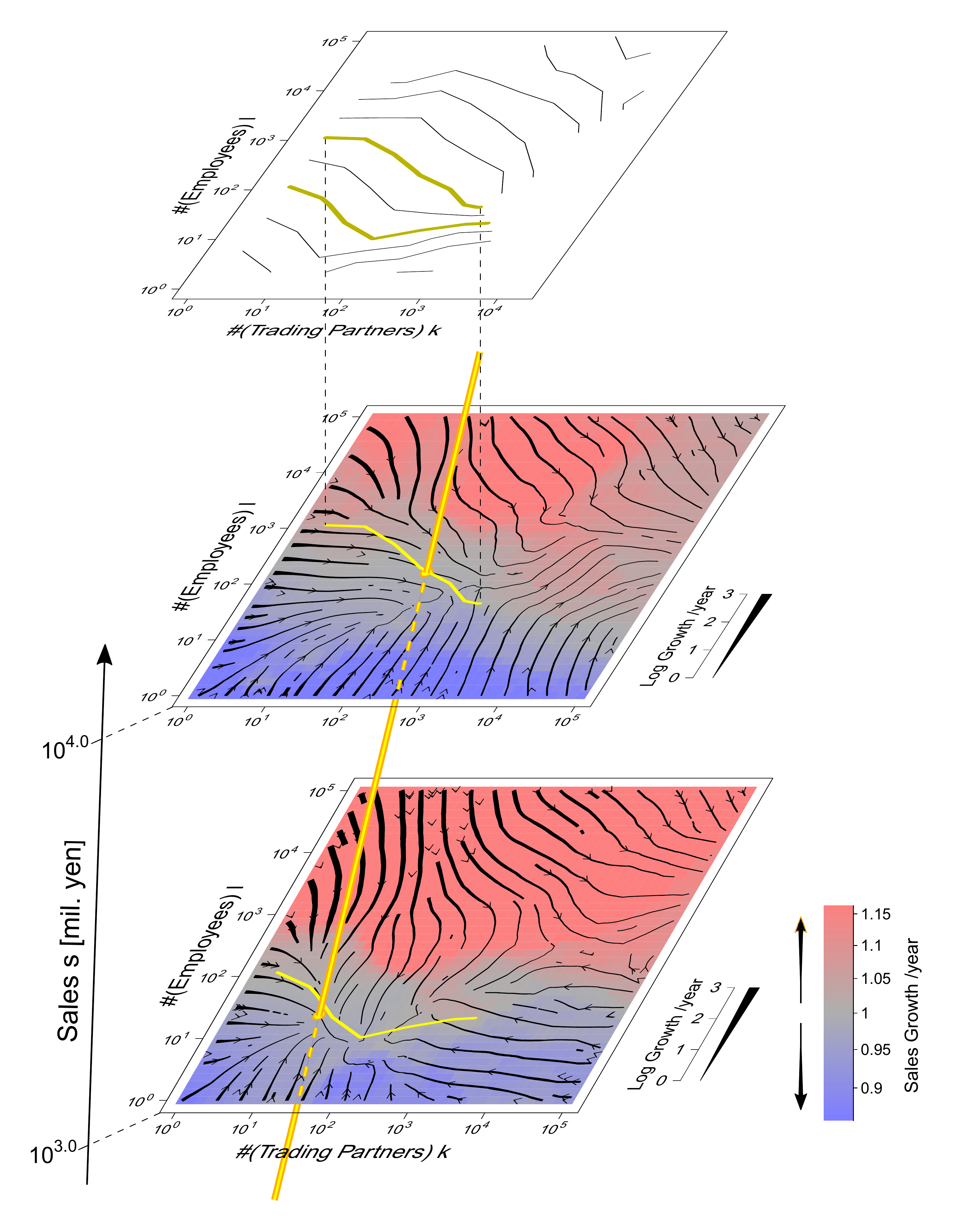}
 }%
 \caption{\label{fig:6} Evolutionary Flow Diagram illustrating the estimated average log-transformed growth per year in slices of the vector space. The variable space is sliced by the plane of annual sales $s$ equal to $10^{3.0}$ or $10^{4.0}$ million yen. The direction and width of the curves in the slices indicate the velocity vector of average flow within the slicing plane. The estimated average flow orthogonal to the plane is illustrated with a background color, red, grey or blue representing plus, zero or minus sales growth, respectively. The orange and yellow line that crosses the slices represents the scaling line of bivariate relations, while yellow curves on the slices indicate the surface of multi-variate scaling, as implied by the contour plot of conditional median sales depending on the numbers of trading partners and employees, placed above the slices. All the plots are obtained based on the total aggregation of data of all years. Note that our method interpolates the average rates so that we have estimation for points around which no firms actually exist.}
\end{figure*}

Considering the robust scaling relations that persists throughout more than 20 years, it is natural to hypothesize that firms that are distant from the scaling surface (Figs.~\ref{fig:2a} and \ref{fig:3b}) have the tendency of flow towards the surface. To validate this hypothesis, we elaborate `evolutionary flow diagrams' by plotting the estimated vector field of annual growth in the three dimensional phase space of $k$, $\ell$ and $s$. This idea is inspired by previous work on the prediction of countries' economic growth\cite{Cristelli2015}, where the authors advocate the applicability of Lorenz's `methods of analogues'\cite{Lorenz1969a,Lorenz1969}, originally proposed for weather forecast, to economic systems. We show some slices of the vector space in Fig.~\ref{fig:6} (also see Supplementary Fig.~\ref{fig:s9}). The streamlines with arrows represent the average movement of firms parallel to the slice, while background colors indicate the average flow of firms orthogonal to the plane. The mean value of log-transformed growth is used: for example, the mean yearly growth of sales is indicated by the average of ${\rm log}[s(t+1)/s(t)]$, where $s(t)$ is the annual sales at year $t$.

One can see the mean flows in sales (background colors of slices in Fig.~\ref{fig:6}) towards the scaling relation surface. Two slices of constant sales ($s = 10^{3.0}$ or $s = 10^{4.0}$ million yen, respectively) are shown in the figure. The intersection curves of the surface and slicing planes are indicated by the yellow curves. Since these contour curves indicate the firm body sizes that yield a specific value of sales for `average' firms (i.e.\ those with median sales for their body sizes), firms in the `back' of the contours in Fig.~\ref{fig:6}, located in large-$\ell$ regions in the slice, have less sales compared to the average firms. Sales of these firms are, therefore, below the average level. They then have positive average growth of sales represented by red background colors in Fig.~\ref{fig:6}, as hypothesized. Conversely, firms with lower $\ell$ below the contour curves in the constant-$s$ slice, which have an excess of sales compared to the average, are very likely to have negative sales growth on average.

Deviations from the scaling surface are compensated not just by sales growth or decrease illustrated vertically to the slice, but also by the move along the slice, i.e.\ their simultaneous changes in body sizes. In fact, firms with sales disproportionate to their body sizes, which are distant from the yellow curves of scaling surface contours in Fig.~\ref{fig:6}, commonly return to the curves (Fig.~\ref{fig:6}), adjusting their body sizes to the current activity rate in sales. Note that the estimates are not so accurate at regions of large body sizes (top-right in Fig.~\ref{fig:6}) or at those with imbalanced configuration (top-left or bottom-right) as for regions of small body sizes (bottom-left), because of poor statistics due to fewer numbers of sample firms. Besides, the continuous increase of trading link data (Supplementary Text \ref{sec:s1}) is likely to add positive bias in the estimate of $k$ flow (the change in number of trading partners), turning the direction of average flows rightward in Fig.~\ref{fig:6}.

\begin{figure*}[t]
 {
  \includegraphics[keepaspectratio, scale=0.43]{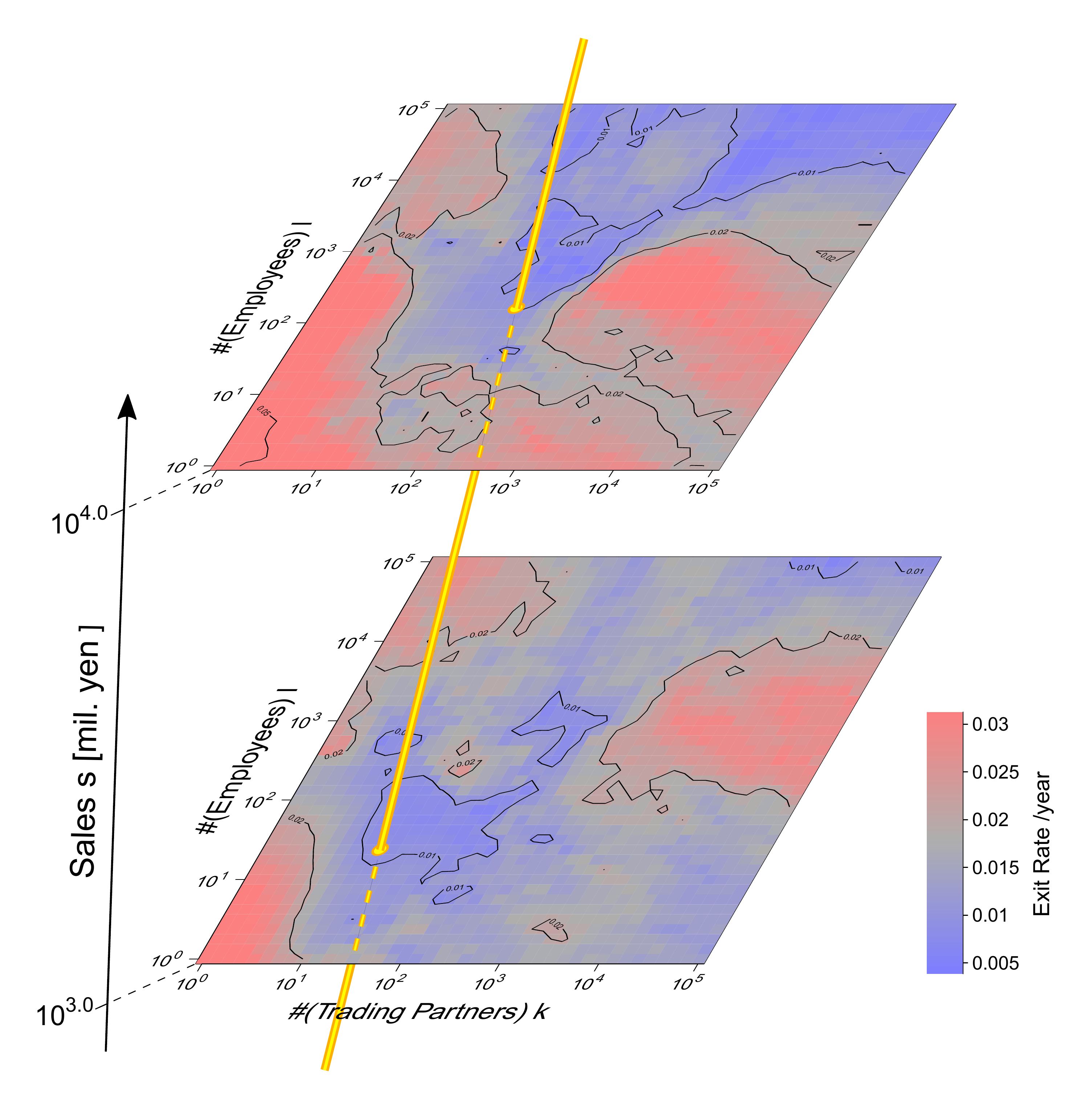}
 }%
 \caption{\label{fig:7} Illustration of the estimated exit rate per year in slices of the vector space. The variable space is sliced by planes of annual sales $s$ equal to $10^{3.0}$ and $10^{4.0}$ million yen. The estimated exit rate is indicated by the background color, red, grey or blue representing high, medium or low rate of exit, respectively. The orange and yellow line that crosses the slices represents the scaling line of bivariate relations. We add contours to show clearly the regions where exit rates are high or low. The plot is obtained based on the total aggregation of data of all years. Note that our method interpolates the average rates so that we have estimation for points around which no firms actually exists.}
\end{figure*}

Consistent with the multi-variate scaling, higher sales growth is expected for firms with more employees, $\ell$, when the initial sales s and number of trading partners $k$ are controlled (Fig.~\ref{fig:6}). On the other hand, the effect of increasing the initial number of trading partners, $k$, on the sales growth is not so visible in the figure. This might be expected from our result above, because $\alpha < \beta$ implies that the gap between the actual sales and the scaled or `balanced' one is larger for firms with more employees rather than for those with more trading partners.

We also notice that those points on the scaling line (Fig.~\ref{fig:3a};\ orange line in Fig.~\ref{fig:6}) are marked with relatively very slow absolute changes in the activity rate and body sizes: average flows around the point of scaling line are close to zero in comparison to other regions of the variable space (see Supplementary Text \ref{sec:s6} for general cases). Therefore, we expect that the growth of firms on the scaling line should be predominantly determined by growth fluctuations and cannot be attributed to their body or activity sizes.

Although we find no direct relation of exit rate to the surface of multi-variate scaling, the scaling line seems to be relevant also to the exit rate of firms, namely, the rate of bankruptcy, merger and suspension or closure of business. For firms of medium or large size, the exit rate exhibits a clear decline around the scaling point (Fig.~\ref{fig:7};\ also see Supplementary Fig.~\ref{fig:s10}). Exit rates are often relatively high for firms that are distant from the scaling line, and at some regions the rates are significantly high, exceeding 3 per cent per year. On the other hand, they are quite low (less than 1 per cent per year) for firms around the scaling line. Thus, deviation from the scaling relations is probably a good sign of higher risk of death.

\section{Discussion}

We have analyzed the scaling relations inherent in firms and their implications on firm dynamics, highlighting firms' general tendency towards scaled states in the 3-dimensional space. We first show that firms are densely located in a 2-dimensional scaling surface, and then demonstrate that there exist evolutionary tendencies that leads firms to the surface. We find that the scaling surface is characterized by a clear asymmetry between the slopes of $s$ (sales) versus $k$ (trading partners) or versus $l$ (employees), where the latter is higher. In other words, the number of employees $\ell$ is more influential to the annual sales $s$ compared to the number of trading partners $k$. If these quantities are not on the scaling surface, they are, on average, adjusted towards their more `balanced' proportions that abide by the scaling relations. Imbalanced firms also have higher tendency to disappear. This means that the scaling relations are maintained dynamically and would be recovered if they were perturbed. It also follows that only the firms deviating from the scaling relations have more chance of higher sales growth, but they also suffer from higher risk for disappearance. This matches the intuitive trade-off between risk and return, whereby one cannot avoid taking higher risk when aspiring to attain higher growth (e.g.\ by increasing recruitment). Of course, this has only partly to do with the whole reality of firms, as random fluctuations in dynamics are prevalent and their increasing employee number does not guarantee positive sales growth, but only increase its chance as evidenced in Fig.~\ref{fig:5}. The results could be directly applied to the prediction of future firm size, which might benefit investors. Another exciting arena of application might be the control of firm development, such as determination of the growth path that maximizes a firm's sales growth for a given risk of disappearance that is maximally bearable for entrepreneurs and other stakeholders.

Note that the overall average flow of firms to a more balanced state on the scaling relation surface does not mean that the firm size distribution eventually reduces to a two-dimensional surface or even a one-dimensional curve. There are always temporal fluctuations in firms' activity rates or sizes. They are the dominant factor of their dynamics especially around the scaling surface (e.g.\ Fig.~\ref{fig:5}), and furthermore, distributions of these `noises' are probably fat-tailed, as the plot of size growth rates suggest (see Supplementary Fig.~\ref{fig:s3}). We speculate that diffusion effects of the stochastic growth rates is in equilibrium with the average flows we just find, leading to the unchanging fat-tailed distribution of firms around the scaling relations (Figs.~\ref{fig:2}b--f) through a process similar to a random multiplicative process\cite{Takayasu1997}. However, the connection between the common scaling function and the stochastic dynamics is yet to be established.

Although we aggregate the data of different years in the evolutionary flow diagrams for the sake of large sample sizes, we find only some small variations when data of different years are compared. One of our important finding is the values of scaling exponents $\alpha$ and $\beta$ and their variations which seemingly follow the country's GDP (Fig.~\ref{fig:4}). The employee number becomes more influential in determining sales in a recovered economy, and the number of trading partners is affecting more (though less than the employee number) in an economic recession compared to other periods. In fact, it is qualitatively convincing that selling whatever produced with labor force would be relatively easy in a recovering economy, while the trading partners to which they could sell their products are more crucial in depression.

We expect that similar results would emerge when applying our method on different datasets of firms from other areas or countries. Tests of this hypothesis would be highly valuable for our understanding of diversity and universality of the firms system. While two-dimensional analysis could be performed without much effort because of the abundant information on the sales and employees, it would be more difficult to conduct a 3-dimensional study of firms in areas other than Japan, since the trading data are often missing.

Our generalized picture of firm dynamics could serve as a possible guide to a unified understanding of many existing results. For example, it was shown that sales growth become higher for firms just after merging, compared to non-merged firms, and the effect slows down with years\cite{Lockett2011}. This could be explained as follows. Assume a situation that the employee number and annual sales of a newly merged firm is the sum of those of the antecedents, and that the antecedents were perfectly on the scaling. Then the annual sales should grow on average, since the annual sales is under the level of scaling, given nonlinear increase of sales against employee increment ($sales \propto employee^{1.2}$ found empirically). Similarly, higher average growth of entering firms\cite{Geurts2016,Gao2016} might be explained with their initial out-scaling relation between their sales and size in employees or trading partners. Thus, generally, the relevance of scaling relations to dynamics found here could explain many features of firm growth. Moreover, the upregulation of company sales after merging is reminiscent of the fact that the metabolic rate per unit mass of a mammalian cell is considerably upregulated when it is cultured \textit{in vitro} with the size of cell clusters far smaller than a mammalian individual\cite{West2002}. Therefore, the general scaling framework developed here could be useful also for understanding other natural or technological complex organizations.

Presenting novel stylized facts, we believe that our results are also beneficial for future modeling and theory construction. Researchers have formulated numerous models\cite{Buldyrev1997,Bottazzi2003,Alfarano2012,Gallegati2003,Wright2005,Fu2005,Riccaboni2008,Mondani2014,Takayasu2014,Malcai1999,Sutton2002,Wyart2003} to explain a few stylized facts on firms and new criteria were apparently needed to discriminate and validate the models. Thus, we suggest that future theoretical studies should incorporate the phenomenological multi-variate evolution of firm entities found here. Theories for scaling relations between sizes in other complex systems such as animal bodies and cities might be relevant to this enterprise, because fractal-like hierarchical organization is a pervasive design in all these systems\cite{Li2015a,West1997,Batty1994,Fix2017}. This might open a prospect of devising general understandings and modeling principles for such complex systems.

\section{Methods}
\subsection{Estimating Scaling Exponents}

We perform the standard regression analysis with R (ver.\ 3.1.2)\cite{RCoreTeam2014} in order to estimate the scaling exponents in the bivariate and multi-variate scaling relationships from the firm data. Bivariate scaling is simply formulated as $x \propto y^{\gamma}$ (defined in a way similar to Eq.\ (2); see Supplementary Text \ref{sec:s2}), where $\gamma$ is the exponent, and $x$ and $y$ are a pair from those three quantities: the number of trading partners, $k$, the number of employees, $\ell$, and annual sales in million yen, $s$. On the other hand, multi-variate one is $s \propto k^{\alpha} \ell^{\beta}$ (Eq.\ (2) for definition), where $\alpha$ and $\beta$ are the scaling exponents of $k$ and $\ell$. Although $k$ and $\ell$ are not mutually independent, regression of $s$ against an orthogonalized set of variables, such as $k$ and $\ell/k^{1.0}$, yields $s \propto k^{\alpha^{\prime}} (\ell/k^{1.0})^{\beta^{\prime}}$, which is equivalent to $s \propto k^{\alpha} \ell^{\beta}$, where $\alpha = {{\alpha}^{\prime}}^{-1.0} \beta^{\prime}$ and $\beta = \beta^{\prime}$ (see Supplementary Text \ref{sec:s4} for more discussion). All probability distribution functions of size variables ($k$, $\ell$ and $s$) are fat-tailed for large values in any year (Supplementary Text \ref{sec:s1}, Supplementary Fig.~\ref{fig:s2}; also see Supplementary Table \ref{tab:s1}). To avoid extreme values usually seen in variables distributed in such a way, we log-transform the raw size figures, so that the variables are exponentially distributed. After the transformation, the model is linear as defined in Eq.\ (2). Although the error terms are distributed in a non-Gaussian manner, they are generally invariable regardless of the value of `explanatory' variables (Fig.~\ref{fig:2a} and Supplementary Figs.~\ref{fig:s4}d--f), and the effect is seemingly linear in larger firms (Figs.~\ref{fig:2}a--c and \ref{fig:4}), so the assumptions of the model Eq.\ (2) are met. We estimate the exponents for every year of 1994--2015. We exclude the data of firms that lack any of the three variables.

To reject the data of small firms that do not fit to the `linear' assumption of the model (Eq.\ (2); also see Eqs.\ (S2.4--6) in Supplementary Text \ref{sec:s2}), we exclude the data with small $k$ or $\ell$ by the threshold of 100. This threshold is determined with regard to consistency of the resulting multi-variate scaling exponents to bivariate ones (Supplementary Texts \ref{sec:s3} and \ref{sec:s4}). Although a considerable fraction of data is missed from the analysis (see Supplementary Fig.~\ref{fig:s5}d for final sample sizes), this makes sure that the resulting set of exponents conforms to the model assumptions.

We determine the confidence intervals for the estimates of $\alpha$ and $\beta$ with the bootstrap technique\cite{Davison1997}. Resampling is performed 10,000 times, with the size of resampling being identical to the sample size. Then 95\% confidence intervals (CI) are estimated with the 2.5- and 97.5-percentiles of the bootstrap distribution.

\subsection{Differentiation of Growth Correlations}

We aim to discriminate between the effects of employee growth and growth in trading partnership on the sales growth. To this end, we compare the firms with a large growth in trading partnership (Group 1) and in employee number (Group 2). `Large growth' is here defined by a growth rate higher than 50\%, and the growth of the other variable is controlled within $\pm 20\%$ to expel the effect of correlation between employee and trading partnership growth from the analysis. Sales growth in year $t$ is defined by the ratio of sales $s$ in year $t$ to that in the initial year $t-1$: i.e.\ $s(t)/s(t-1)$. We consider the `accompanying' sales growth in year $t$, as well as the `following' sales growth, in year $t+1$. Then, we apply the two-sample Kolmogorov-Smirnov test to the two empirical distributions of sales growth rates from both groups with R (ver.\ 3.1.2)\cite{RCoreTeam2014}. Two-tailed test is performed with the significance level of 0.05. Nonparametric tests are favored here, since the distribution is possibly non-Gaussian (Fig.~\ref{fig:5}) and there is not unanimous agreement on which family of distributions should be fitted against the empirical growth rates\cite{Williams2015}.

To rule out the possibility that size heterogeneity affects the results, the size variables in the initial year $t-1$ are also controlled. Let us define $d_{\rm log}$ as a firm's Euclidean distance in the logarithmically scaled space from a fixed point $(k_0, \ell_0, s_0)$ in the initial year:
\begin{widetext}
\begin{equation}
d_{\rm log} =
\sqrt{
\left( {\rm log}[k(t-1)/k_0] \right)^2 +
\left( {\rm log}[\ell(t-1)/\ell_0] \right)^2 +
\left( {\rm log}[s(t-1)/s_0] \right)^2
}.
\end{equation}
\end{widetext}
We include only the firms within the Euclid radius of $d_{\rm log} < {\rm log}[10^{1/8}]$. The fixed point of initial sizes is set on the `scaling line' (Fig.~\ref{fig:3a};\ for the definition, see Eq.\ (S6.1) in Supplementary Text \ref{sec:s6}), to approximately maximize the sample density around the point and to avoid possible biases. A satisfactory sample size is assured by aggregating the whole data of all years, for which we estimate the scaling exponents $\alpha$ and $\beta$ as in the above section. Data around the scaling point of $k = 10$ is used to produce the illustrative results in the main text; we address the dependence of results on the $k$ value choice in Supplementary Text \ref{sec:s5} (also see Supplementary Fig.~\ref{fig:s7}).

Additionally, we calculate the mean value of log-transformed accompanying or following sales growth rates in these groups. We determine the confidence intervals of the mean again with the bootstrap technique. Here, we apply the same procedures as described in the above section to have 95\% CI. We loosely use the term `significance' of difference when no overlap exists between two 95\% CI.

\subsection{Estimating Growth and Exit Rates}

Here we estimate the medium growth rates of size variables, $k$, $\ell$ and $s$, at a specific point $(k_0, \ell_0, s_0)$ in the 3-dimensional vector space in order to draw the evolutionary phase diagrams (Fig.~\ref{fig:6}). We do this again by collecting the data sufficiently near the point and computing the arithmetic mean of log-transformed growth rates. We sample the firms of $d_{\rm log} < {\rm log}[10^{1/4}]$, eventually getting Nadaraya-Watson estimate with the kernel function of rectangular pulse\cite{Hastie2009}. Nevertheless, when the resulting sample size is less than 200, the threshold of $d_{\rm log}$ is enlarged until the sample number exceeds 200 to suppress the variability of estimates, therefore employing the 200-nearest neighbor method. Taking the logarithm of growth rates, we can limit their possible ranges of several order of magnitude (Supplementary Text \ref{sec:s1}, Supplementary Fig.~\ref{fig:s3}) within those expected from exponential distributions, which makes the arithmetic mean a more robust estimator of the typical value. Note that the arithmetic mean of log-transformed growth rates is equal to the logarithm of geometric mean of growth rates.

We use almost the same method for exit rate estimations (Fig.~\ref{fig:7}), except that the threshold of the sample size is 1,000 rather than 200. This choice is due to the generally low rate of firm exit, only up to 0.03 per year and sometimes lower than 0.01 per year.

Note that the growth rates could be estimated for points around which almost no firms actually exist. In such a case, these estimates are based on the firms on the nearest edge of distribution. Also, be cautious that they are biased when there is gradient of data density: the center of the distribution then has more weight than peripheral regions, so that the effects of moving outward from the center of distribution on growth rate changes are always underestimated.

\subsection{Code Availability}

Source codes used in this study is available upon request to the corresponding author, which are written in R language (ver.\ 3.1.2)\cite{RCoreTeam2014} and in ShellScript compatible with Red Hat Enterprise Linux Workstation release 7.0 (Red Hat,\ Raleigh,\ NC,\ USA).

\subsection{Data Availability}

The data that support the findings of this study are available from Teikoku Databank,\ Ltd., Japan, but restrictions apply to the availability of these data, which were used under license for the current study, and so are not publicly available. Data are however available from the authors upon reasonable request and with permission of Teikoku Databank,\ Ltd.

\begin{acknowledgments}
We thank O.\ Levy for discussions. The authors appreciate Teikoku Databank,\ Ltd., Center for TDB Advanced Data Analysis and Modeling for providing both the data and financial support. This work is partially supported by the Grant-in-Aid for Scientific Research (B), Grant Number 26310207 and JST, Strategic International Collaborative Research Program (SICORP) on the topic of ``ICT for a Resilient Society'' by Japan and Israel, and by MEXT as ``Exploratory Challenges on Post-K computer (Study on multilayered multiscale spacetime simulations for social and economical phenomena)''.
\end{acknowledgments}

\renewcommand\acknowledgmentsname{Author Contributions}
\begin{acknowledgments}
M.T.\ directed the project. Y.K., H.T.\ and S.H.\ developed the data analysis. Y.K.\ analyzed the data and generated the diagrams. All authors contributed in writing the paper.
\end{acknowledgments}

\renewcommand\acknowledgmentsname{Competing Interests}
\begin{acknowledgments}
Teikoku Databank,\ Ltd.\ supported our research by providing the data regarding Japanese business firms and by financially supporting Center for TDB Advanced Data Analysis and Modeling,\ Tokyo Institute of Technology for academic research purposes. Teikoku Databank,\ Ltd.\ did not participate in our research or preparation of the manuscript except the data collection.
\end{acknowledgments}

\bibliography{kobayashi}% Produces the bibliography via BibTeX.

%\includepdf[pages=1,lastpage=10]{SI.pdf}

\onecolumngrid

\newpage
\renewcommand\appendixname{Supplementary Text}
\appendix
\setcounter{figure}{0}
\setcounter{section}{0}
\renewcommand{\thesection}{\arabic{section}}
\renewcommand{\thefigure}{S\arabic{figure}}
%\capnumoff=1

\makeatletter
\renewcommand\fnum@figure{}
\renewcommand\figurename{}

\makeatother

\vspace*{5pt}
\centering
\includegraphics[keepaspectratio, scale=1.05, page=1, trim=23.5mm 38.5mm 23.5mm 38.5mm, clip]{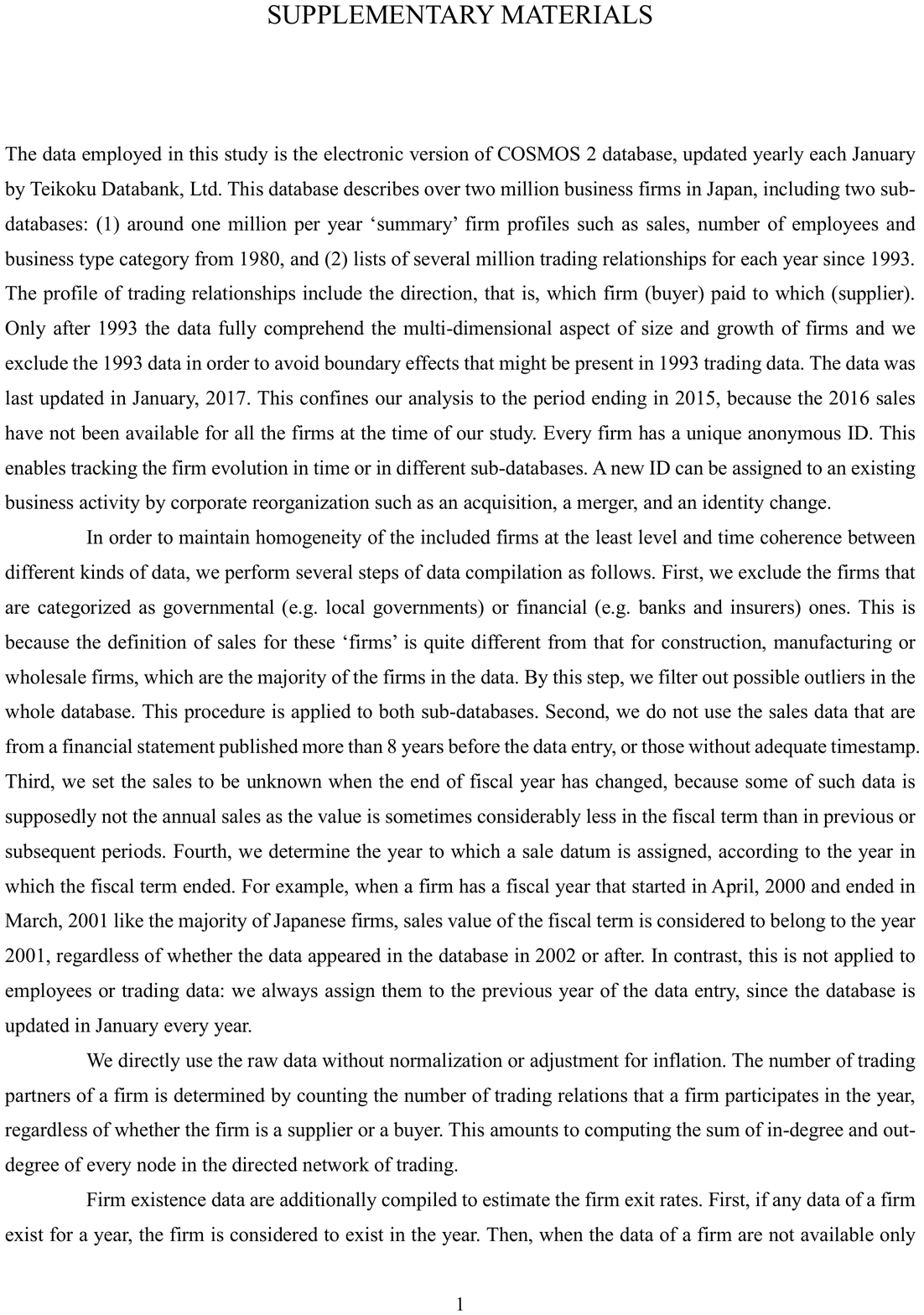}

\vspace{-22.25cm}

\section{\label{sec:s1} Data Description}

\setcounter{page}{1}
\renewcommand{\thepage}{S\arabic{page}}

\newpage
\vspace*{5pt}
\centering
\includegraphics[keepaspectratio, scale=1.05, page=2, trim=23.5mm 38.5mm 23.5mm 38.5mm, clip]{SI.pdf}

\newpage
\vspace*{5pt}
\centering
\includegraphics[keepaspectratio, scale=1.05, page=3, trim=23.5mm 38.5mm 23.5mm 38.5mm, clip]{SI.pdf}

\vspace{-24.25cm}

\section{\label{sec:s2} Scaling between Variables and Universal Distributions}

\vspace{21.25cm}

\section{\label{sec:s3} Derivation of Asymptotic Power-law or Scaling Exponents}

\newpage
\vspace*{5pt}
\centering
\includegraphics[keepaspectratio, scale=1.05, page=4, trim=23.5mm 38.5mm 23.5mm 38.5mm, clip]{SI.pdf}

\newpage
\vspace*{5pt}
\centering
\includegraphics[keepaspectratio, scale=1.05, page=5, trim=23.5mm 38.5mm 23.5mm 38.5mm, clip]{SI.pdf}

\newpage
\vspace*{5pt}
\centering
\includegraphics[keepaspectratio, scale=1.05, page=6, trim=23.5mm 38.5mm 23.5mm 38.5mm, clip]{SI.pdf}

\newpage
\vspace*{5pt}
\centering
\includegraphics[keepaspectratio, scale=1.05, page=7, trim=23.5mm 38.5mm 23.5mm 38.5mm, clip]{SI.pdf}

\newpage
\vspace*{5pt}
\centering
\includegraphics[keepaspectratio, scale=1.05, page=8, trim=23.5mm 38.5mm 23.5mm 38.5mm, clip]{SI.pdf}

\newpage
\vspace*{5pt}
\centering
\includegraphics[keepaspectratio, scale=1.05, page=9, trim=23.5mm 38.5mm 23.5mm 38.5mm, clip]{SI.pdf}

\newpage
\vspace*{5pt}
\centering
\includegraphics[keepaspectratio, scale=1.05, page=10, trim=23.5mm 38.5mm 23.5mm 38.5mm, clip]{SI.pdf}

\vspace{-24.25cm}

\section{\label{sec:s4} Estimating Scaling Exponents}

\newpage
\vspace*{5pt}
\centering
\includegraphics[keepaspectratio, scale=1.05, page=11, trim=23.5mm 38.5mm 23.5mm 38.5mm, clip]{SI.pdf}

\newpage
\vspace*{5pt}
\centering
\includegraphics[keepaspectratio, scale=1.05, page=12, trim=23.5mm 38.5mm 23.5mm 38.5mm, clip]{SI.pdf}

\newpage
\vspace*{5pt}
\centering
\includegraphics[keepaspectratio, scale=1.05, page=13, trim=23.5mm 38.5mm 23.5mm 38.5mm, clip]{SI.pdf}

\vspace{-7.25cm}

\section{\label{sec:s5} Generality of Growth Correlations}

\newpage
\vspace*{5pt}
\centering
\includegraphics[keepaspectratio, scale=1.05, page=14, trim=23.5mm 38.5mm 23.5mm 38.5mm, clip]{SI.pdf}

\newpage
\vspace*{5pt}
\centering
\includegraphics[keepaspectratio, scale=1.05, page=15, trim=23.5mm 38.5mm 23.5mm 38.5mm, clip]{SI.pdf}

\vspace{-22.25cm}

\section{\label{sec:s6} Evolutionary Flow Diagram}

\newpage
\vspace*{5pt}
\centering
\includegraphics[keepaspectratio, scale=1.05, page=16, trim=23.5mm 38.5mm 23.5mm 38.5mm, clip]{SI.pdf}

\newpage
\vspace*{5pt}
\centering
\includegraphics[keepaspectratio, scale=1.05, page=17, trim=23.5mm 38.5mm 23.5mm 38.5mm, clip]{SI.pdf}

\newpage
\vspace*{5pt}
\centering
\includegraphics[keepaspectratio, scale=1.05, page=18, trim=23.5mm 38.5mm 23.5mm 38.5mm, clip]{SI.pdf}

\vspace*{-25cm}

\begin{figure*}[h]
 \includegraphics[keepaspectratio]{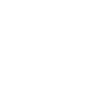}
 \caption{\label{fig:s1}}
\end{figure*}

\newpage
\vspace*{5pt}
\centering
\includegraphics[keepaspectratio, scale=1.05, page=19, trim=23.5mm 38.5mm 23.5mm 38.5mm, clip]{SI.pdf}

\vspace*{-25cm}

\begin{figure*}[h]
 \includegraphics[keepaspectratio]{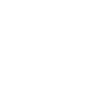}
 \caption{\label{fig:s2}}
\end{figure*}

\newpage
\vspace*{5pt}
\centering
\includegraphics[keepaspectratio, scale=1.05, page=20, trim=23.5mm 38.5mm 23.5mm 38.5mm, clip]{SI.pdf}

\vspace*{-25cm}

\begin{figure*}[h]
 \includegraphics[keepaspectratio]{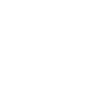}
 \caption{\label{fig:s3}}
\end{figure*}

\newpage
\vspace*{5pt}
\centering
\includegraphics[keepaspectratio, scale=1.05, page=21, trim=23.5mm 38.5mm 23.5mm 38.5mm, clip]{SI.pdf}

\vspace*{-5cm}

\begin{figure*}[h]
 \includegraphics[keepaspectratio]{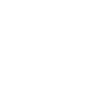}
 \caption{\label{fig:s4}}
\end{figure*}

\newpage
\vspace*{5pt}
\centering
\includegraphics[keepaspectratio, scale=1.05, page=22, trim=23.5mm 38.5mm 23.5mm 38.5mm, clip]{SI.pdf}

\newpage
\vspace*{5pt}
\centering
\includegraphics[keepaspectratio, scale=1.05, page=23, trim=23.5mm 38.5mm 23.5mm 38.5mm, clip]{SI.pdf}

\vspace*{-25cm}

\begin{figure*}[h]
 \includegraphics[keepaspectratio]{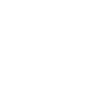}
 \caption{\label{fig:s5}}
\end{figure*}

\newpage
\vspace*{5pt}
\centering
\includegraphics[keepaspectratio, scale=1.05, page=24, trim=23.5mm 38.5mm 23.5mm 38.5mm, clip]{SI.pdf}

\vspace*{-25cm}

\begin{figure*}[h]
 \includegraphics[keepaspectratio]{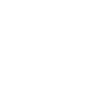}
 \caption{\label{fig:s6}}
\end{figure*}

\newpage
\vspace*{5pt}
\centering
\includegraphics[keepaspectratio, scale=1.05, page=25, trim=23.5mm 38.5mm 23.5mm 38.5mm, clip]{SI.pdf}

\vspace*{-25cm}

\begin{figure*}[h]
 \includegraphics[keepaspectratio]{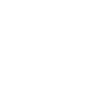}
 \caption{\label{fig:s7}}
\end{figure*}

\newpage
\vspace*{5pt}
\centering
\includegraphics[keepaspectratio, scale=1.05, page=26, trim=23.5mm 38.5mm 23.5mm 38.5mm, clip]{SI.pdf}

\vspace*{-25cm}

\begin{figure*}[h]
 \includegraphics[keepaspectratio]{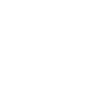}
 \caption{\label{fig:s8}}
\end{figure*}

\newpage
\vspace*{5pt}
\centering
\includegraphics[keepaspectratio, scale=1.05, page=27, trim=23.5mm 38.5mm 23.5mm 38.5mm, clip]{SI.pdf}

\vspace*{-25cm}

\begin{figure*}[h]
 \includegraphics[keepaspectratio]{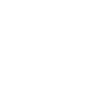}
 \caption{\label{fig:s9}}
\end{figure*}

\newpage
\vspace*{5pt}
\centering
\includegraphics[keepaspectratio, scale=1.05, page=28, trim=23.5mm 38.5mm 23.5mm 38.5mm, clip]{SI.pdf}

\newpage
\vspace*{5pt}
\centering
\includegraphics[keepaspectratio, scale=1.05, page=29, trim=23.5mm 38.5mm 23.5mm 38.5mm, clip]{SI.pdf}

\newpage
\vspace*{5pt}
\centering
\includegraphics[keepaspectratio, scale=1.05, page=30, trim=23.5mm 38.5mm 23.5mm 38.5mm, clip]{SI.pdf}

\newpage
\vspace*{5pt}
\centering
\includegraphics[keepaspectratio, scale=1.05, page=31, trim=23.5mm 38.5mm 23.5mm 38.5mm, clip]{SI.pdf}

\vspace*{-25cm}

\begin{figure*}[h]
 \includegraphics[keepaspectratio]{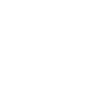}
 \caption{\label{fig:s10}}
\end{figure*}

\newpage
\vspace*{5pt}
\centering
\includegraphics[keepaspectratio, scale=1.05, page=32, trim=23.5mm 38.5mm 23.5mm 38.5mm, clip]{SI.pdf}

\newpage
\vspace*{5pt}
\centering
\includegraphics[keepaspectratio, scale=1.05, page=33, trim=23.5mm 38.5mm 23.5mm 38.5mm, clip]{SI.pdf}

\newpage
\vspace*{5pt}
\centering
\includegraphics[keepaspectratio, scale=1.05, page=34, trim=23.5mm 38.5mm 23.5mm 38.5mm, clip]{SI.pdf}

\newpage
\vspace*{5pt}
\centering
\includegraphics[keepaspectratio, scale=1.02, page=35, trim=23.5mm 38.5mm 23.5mm 37mm, clip]{SI.pdf}

\setcounter{figure}{0}

\vspace*{-25cm}

\begin{figure*}[h]
 \includegraphics[keepaspectratio]{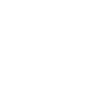}
 \caption{\label{tab:s1}}
\end{figure*}

\newpage
\vspace*{5pt}
\centering
\includegraphics[keepaspectratio, scale=1.02, page=36, trim=23.5mm 38.5mm 23.5mm 37mm, clip]{SI.pdf}

\newpage
\vspace*{5pt}
\centering
\includegraphics[keepaspectratio, scale=1.02, page=37, trim=23.5mm 38.5mm 23.5mm 37mm, clip]{SI.pdf}

\newpage
\vspace*{5pt}
\centering
\includegraphics[keepaspectratio, scale=1.05, page=38, trim=23.5mm 38.5mm 23.5mm 38.5mm, clip]{SI.pdf}

\end{document}